\documentclass{article}

\newcommand {\un}      {\,\mbox}
\newcommand {\GeV}      {\un{GeV}}
\newcommand {\fbo}        {\,\ensuremath{\mathrm {fb}^{-1}}}
\newcommand {\pbo}        {\,\ensuremath{\mathrm {pb}^{-1}}}
\newcommand {\pt}         {\ensuremath{p_T}}
\newcommand {\etajet} {\ensuremath{\eta^{\mathrm {jet}}}}

\newcommand {\bbbar}	{\ensuremath{b\bbar}}
\newcommand {\bbar}	{\ensuremath{\bar b}}
\newcommand {\ccbar}	{\ensuremath{c\cbar}}
\newcommand {\cbar}	{\ensuremath{\bar c}}

\newcommand {\ppbar}	{\ensuremath{p\pbar}}
\newcommand {\pbar}	{\ensuremath{\bar p}}
\newcommand {\ttbar}    {\ensuremath{t\tbar}}
\newcommand {\tbar}     {\ensuremath{\bar t}}
\newcommand {\W}        {\ensuremath{W}}
\newcommand {\Z}        {\ensuremath{Z}}
\newcommand {\V}        {\ensuremath{V}}
\newcommand {\epem}     {\ensuremath{e^+e^-}}
\newcommand {\mpmm}     {\ensuremath{\mu^+\mu^-}}
\newcommand {\tptm}     {\ensuremath{\tau^+\tau^-}}

\newcommand {\wpj}      {\ensuremath{W\hspace{-1pt}\mathrm{+jets}}}
\newcommand {\wpc}      {\ensuremath{W\hspace{-1pt}\mathrm{+}c}}
\newcommand {\wpcj}     {\ensuremath{W+c\mathrm{-jet}}}
\newcommand {\wpbj}     {\ensuremath{W+b\mathrm{-jet}}}
\newcommand {\zpj}      {\Z+jets}
\newcommand {\vpj}      {\V+jets}

\newcommand {\whf}      {\ensuremath{\W+\mathrm{HF}}}
\newcommand {\wlf}      {\ensuremath{\W+\mathrm{LF}}}
\newcommand {\khf}      {\ensuremath{K_\mathrm{HF}}}

\newcommand {\DZ}         {D0} % Changed recently from D\0 for electronic searches (D0 style guidelines 4.0)
\newcommand {\CDF}        {CDF}

\newcommand {\stat} 	{\ensuremath{\left(\mathrm{stat.} \right)}}
\newcommand {\syst} 	{\ensuremath{\left(\mathrm{syst.} \right)}}
\newcommand {\lumi} 	{\ensuremath{\left(\mathrm{lumi.} \right)}}
\newcommand {\pdfs} 	{\ensuremath{\left(\mathrm{PDF} \right)}}
 \newcommand {\etal}     {{\it et al.}}

\newcommand {\alpgen}   {{\sc alpgen}}
\newcommand {\herwig}   {{\sc herwig}}
\newcommand {\pythia}   {{\sc pythia}}
\newcommand {\resbos}   {{\sc resbos}}
\newcommand {\sherpa}   {{\sc sherpa}}
 
\newcommand{\halfWid}	{0.48}

\usepackage{graphicx}  % got figures? uncomment this
\title{Monte Carlo for top background at the Tevatron}
\author{A.~Harel on behalf of the \CDF\ and \DZ\ collaborations.}
\date{Proceedings of talk given at the \\ International Workshop on Top Quark Physics May 22, 2008}
\begin{document}

\maketitle

\begin{abstract}
We review the use of Monte Carlo (MC) simulation 
to model backgrounds to top signal at
the Tevatron experiments, \CDF\ and \DZ, as well
as the relevant measurements done by the experiments.
We'll concentrate on the modeling of \W\ and \Z\ 
boson production in association with jets, in 
particular heavy flavor jets (HF), and also comment
on the Tevatron experience using matched MC.
\end{abstract}

\section {Introduction}

The Fermilab Tevatron Collider has provided over $4\fbo$
of \ppbar\ collisions at $\sqrt{s}=1.96\GeV$, allowing
the \CDF\ and \DZ\ experiments to make precise measurements
using \ttbar\ production, and to find evidence for the
rare single top production process.
Both endeavors require a solid understanding of the
background processes, and MC simulation is a crucial
ingredient of the background models used.

\section {The background processes}

%\subsection{Top pairs}

\begin{figure}
\centering
\mbox{
  \includegraphics[clip, bb=40 0 567 569, width=0.4\linewidth]{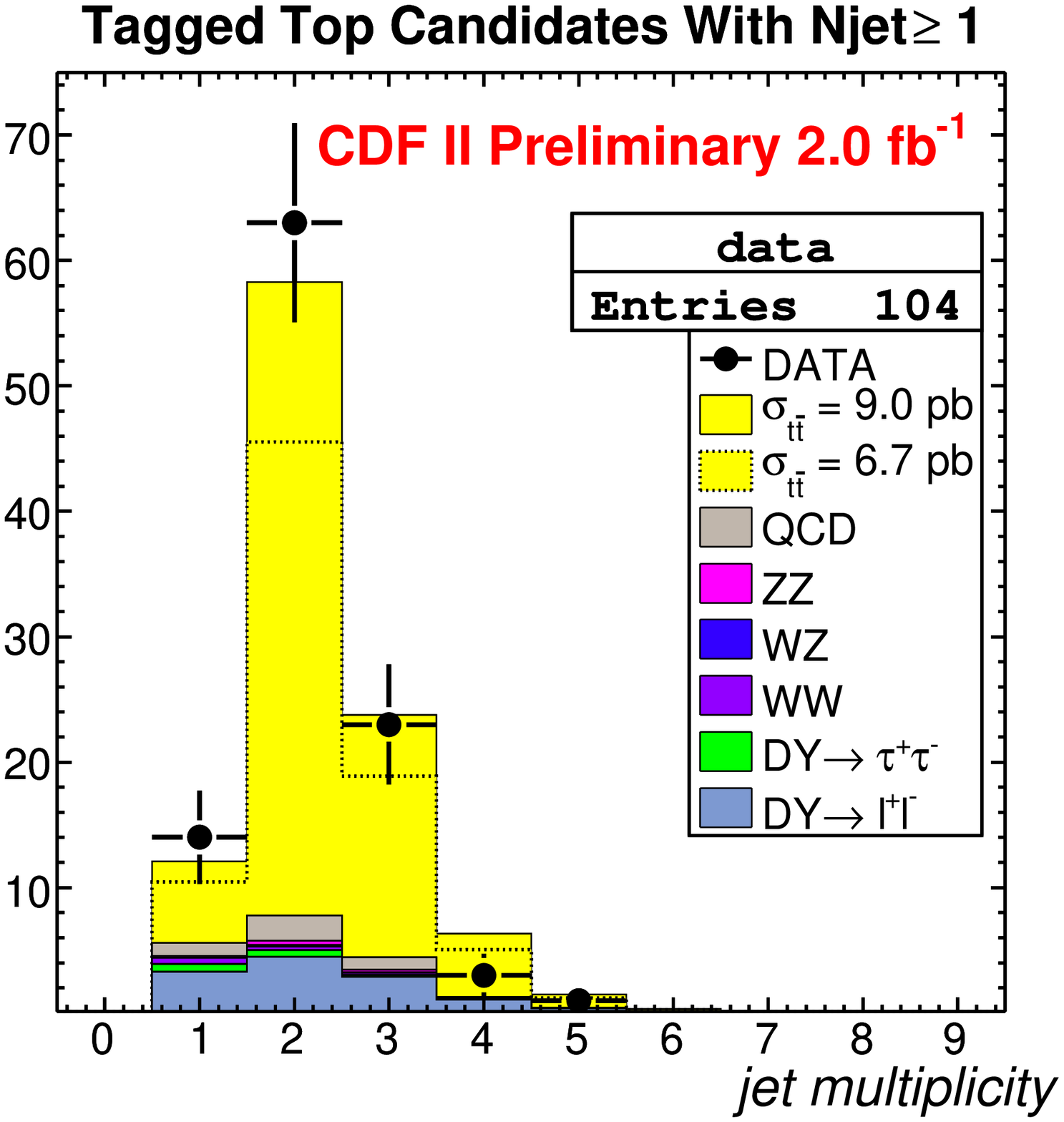}
  \includegraphics[width=0.55\linewidth]{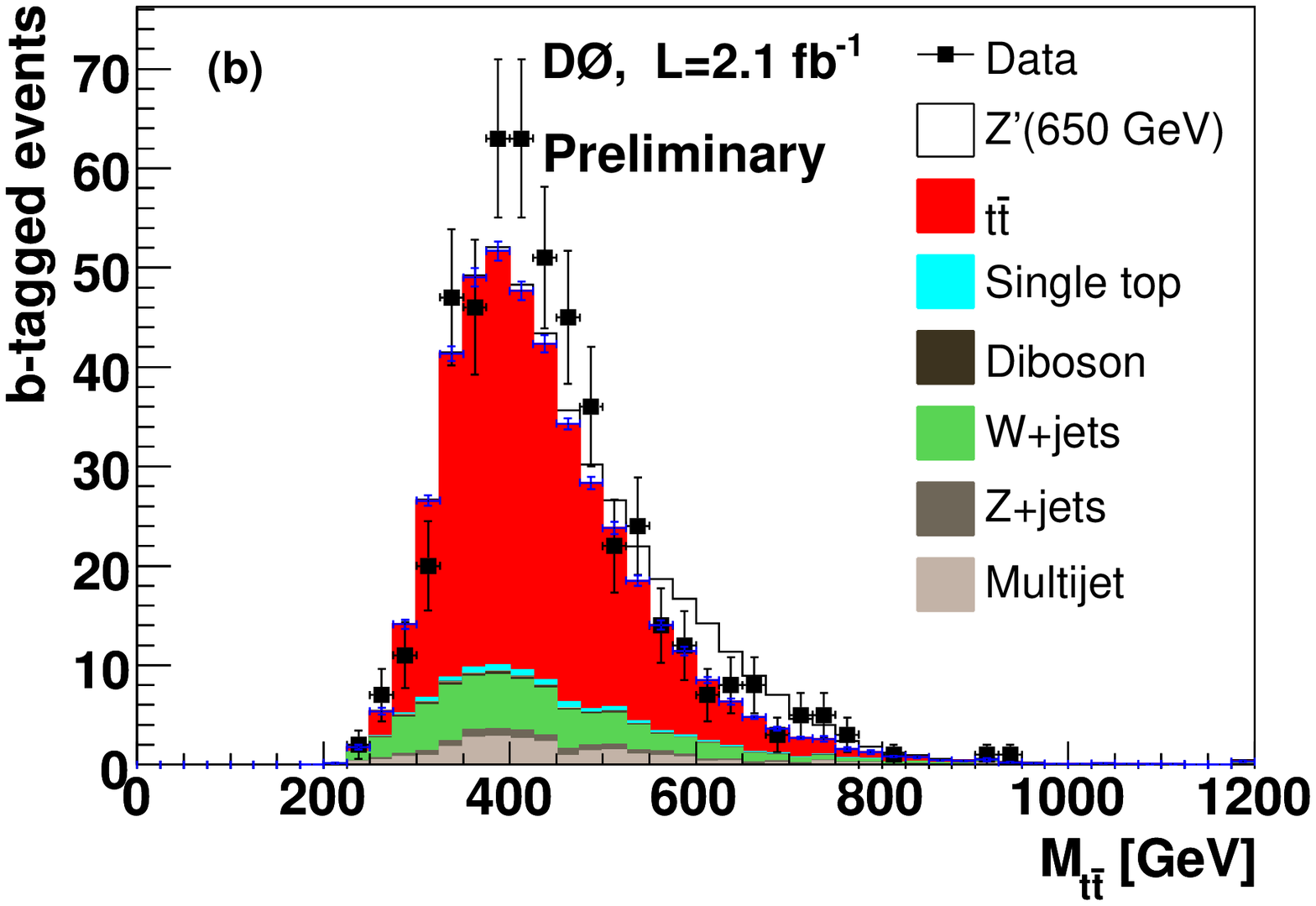} % manually edited for B&W printing
}
\caption{\label{fig:toppair}
Examples of sample composition in top pair analyses for the
dilepton channel~\cite{ref:dilep} (left) and 
the lepton plus jets channel~\cite{ref:ljets} (right).
The ``DY'' label refers Drell-Yan production, as does the ``Z+jets'' label.
The right plot is taken from a search for resonant \ttbar\ production,
and shows (in white) a conceivable new physics component.
%The ``QCD'' label refers to multijet production.
}
\end{figure}

Precision measurements of top quark properties are performed
by studying \ttbar\ production in either:
\begin{itemize}
\item the dilepton decay channel, where the $t \to b\W$ decays are followed 
with $\W \to l\nu(X)$ and $l$ is an electron or muon. Or in
\item the semileptonic (``lepton plus jets'') decay channel, 
where one of the \W\ bosons decays as $\W \to l\nu(X)$
and the other decays hadronically.
\end{itemize}
Fig~\ref{fig:toppair} shows typical sample compositions
 in these channels.

The dominate background in the dilepton channels
is Drell-Yan plus jets production,
$\Z\to\epem$ in the $\epem$ channel, 
$\Z\to\mpmm$ in the $\mpmm$ channel, and
$\Z\to\tptm$ with subsequent leptonic $\tau$ decays in the $e\mu$ channel.
In these proceedings we'll follow the common practice of 
referring to this background as ``\zpj''.
This background dominates the early stages of the
event selection, when the experimental understanding of the
samples is verified, e.g., by examining many distributions
for control samples with fewer jets than the signal.
But after the final selection, its 
contribution is small.
Dilepton samples are quite pure, hence most analyses do not
rely on $b$ tagging and the precise flavor composition of the
\zpj\ background is not important.

The second largest background is multijet production, often
referred to as ``QCD'' background. Multijet events are selected
when jets are misreconstructed as leptons. It is quite
difficult to simulate these mistaken reconstructions both
at the MC generator level and at the detector simulation level.
Therefore data driven models are used for these backgrounds
(see also sec~\ref{sec:multijet}).
The next background component is from diboson plus jets production.
These background are quite small, so even a rough simulation
suffices for top physics, and they are estimated purely from MC.

In the lepton plus jet channel, the dominate background is
\wpj\ production, which is important both in the control samples
and in the signal samples.
This channel provides the most precise measurements, and
most measurements use $b$ tagging to suppress background~\cite{ref:btagintop}.
As a result, the flavor composition of the jets produced in
association with a \W\ boson is relevant in this channel.

\begin{figure}
\centering
\includegraphics[width=\halfWid\linewidth]{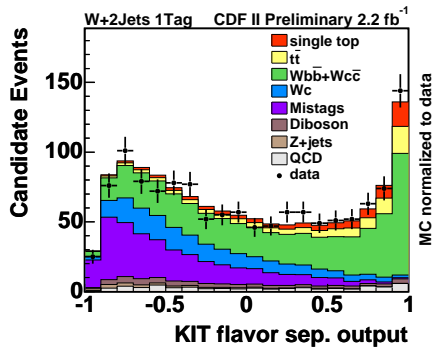} % manually edited for B&W printing
\caption{\label{fig:singletop}
Sample composition in \CDF's single top search.
The ``mistag'' label refers to \W\ plus light-flavored jets
production. % The ``QCD'' label refers to multijet production.
}
\end{figure}

The search for single top production is characterized
by high level of background, as the experimental signature
of this process contains fewer jets. 
Thus the single top analyses use $b$ tagging to
suppress background.
A typical sample composition is 
shown in fig~\ref{fig:singletop}.
These samples are dominated by \wpj\ production,
and knowledge of the flavor composition of the jets produced in
association with a \W\ boson is required to identify the small
single-top signal.

\section {Matched MC in \boldmath \vpj}
\label{sec:match}

The calculation of the differential cross sections for
\wpj\ and \zpj\ processes (\vpj), 
and in particular \W\ plus heavy flavor (\whf)
production, is far from trivial.
It was the motivation for the development of the \alpgen\
event generator~\cite{ref:alpgen1}, and recent calculations
show that sizable NLO corrections
exist for some final states~\cite{ref:wbjcalc}.

Production of hard additional partons is well simulated by
matrix element (ME) generators that calculate $2\to n$ processes
at tree level, such as \alpgen.
But parton shower (PS) MC, such as \pythia~\cite{ref:pythia},
are better at simulating softer radiation,
as the PS approximates the sum of soft
contributions from all orders in perturbation theory.
Hence these tools are used together, the hard $2\to n$ interactions
being modeled by the ME generator, and the showering by
the PS generator. Care must be taken to avoid double
counting final states,  for example, those where the 3rd hardest parton
can be generated either by the ME or by the PS.
This is done using a matching prescription, discussed
elsewhere~\cite{ref:matching}.

The \CDF\ and \DZ\ collaborations both generate \vpj\ 
MC with \alpgen\ using the MLM matching prescription~\cite{ref:MLM},
with some small differences in the matching technology.
Since \whf\ production is important for top physics, 
both collaborations produce such samples separately. 
But these samples overlap with the \wpj\ 
samples, which include heavy flavor jets in the PSs,
and this overlap must be removed.
The \CDF\ collaboration does so	by classifying \bbbar\ and 
\ccbar\ pairs into those that are in the same parton jet
and those that are not. The former are taken only from the
PS MC (\herwig~\cite{ref:herwig}), 
and the latter only from the
ME MC (\alpgen). This has the advantage of
playing to each MC's strength.
The \DZ\ collaboration uses the more straight-forward
solution of discarding any events 
that were generated as \wpj\ by the ME MC (\alpgen) and contain
heavy-flavor jets added by the PS MC (\pythia).

Other differences are in the \pt\ cut used for the matching
within each sample ($15\GeV$ in \CDF, $8\GeV$ in \DZ), which
has little effect, in the light-parton jet multiplicities
produced for each sample (up to 4 in \CDF, up to 5 in \DZ),
and in the treatment of \wpc\ production (separate
in \CDF, included in \wlf\ in \DZ).

\section {Measurements of \boldmath \vpj\ processes}

Given the difficulties in calculating and simulation
\vpj\ processes, it is instructive to compare them
to data. In this section we review measurements of
\vpj\ production from \CDF\ and \DZ. The leptonic \W\ and
\Z\ decay channels provide clear experimental
signatures and are used throughout.
Since the additional jets are produced by the strong interaction,
which favors soft and collinear radiation, selection cuts
on energies and angles have a large effect on the
cross sections.
Relevant selection cuts will be stressed in this section.

\begin{figure}
\begin{minipage}[b]{0.48\linewidth}
  \centering
  \includegraphics[width=\linewidth]{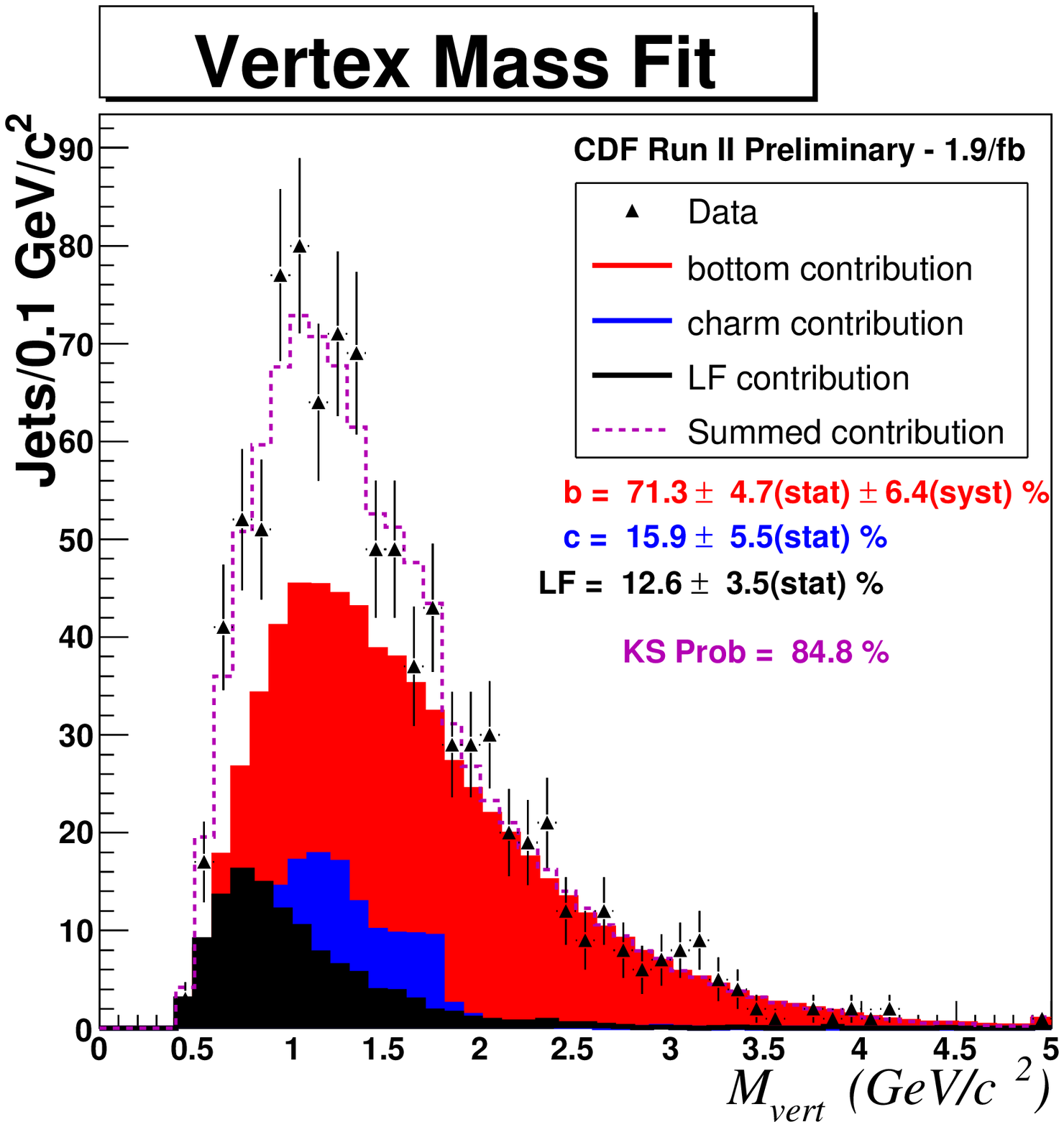}
  \caption{\label{fig:cdfwb}
  Vertex mass fit for tagged jets in selected sample of ref~\cite{ref:cdfwb}.
  }
\end{minipage}
\hspace{0.04\linewidth}
\begin{minipage}[b]{0.48\linewidth}
 \centering
 \includegraphics[width=\linewidth]{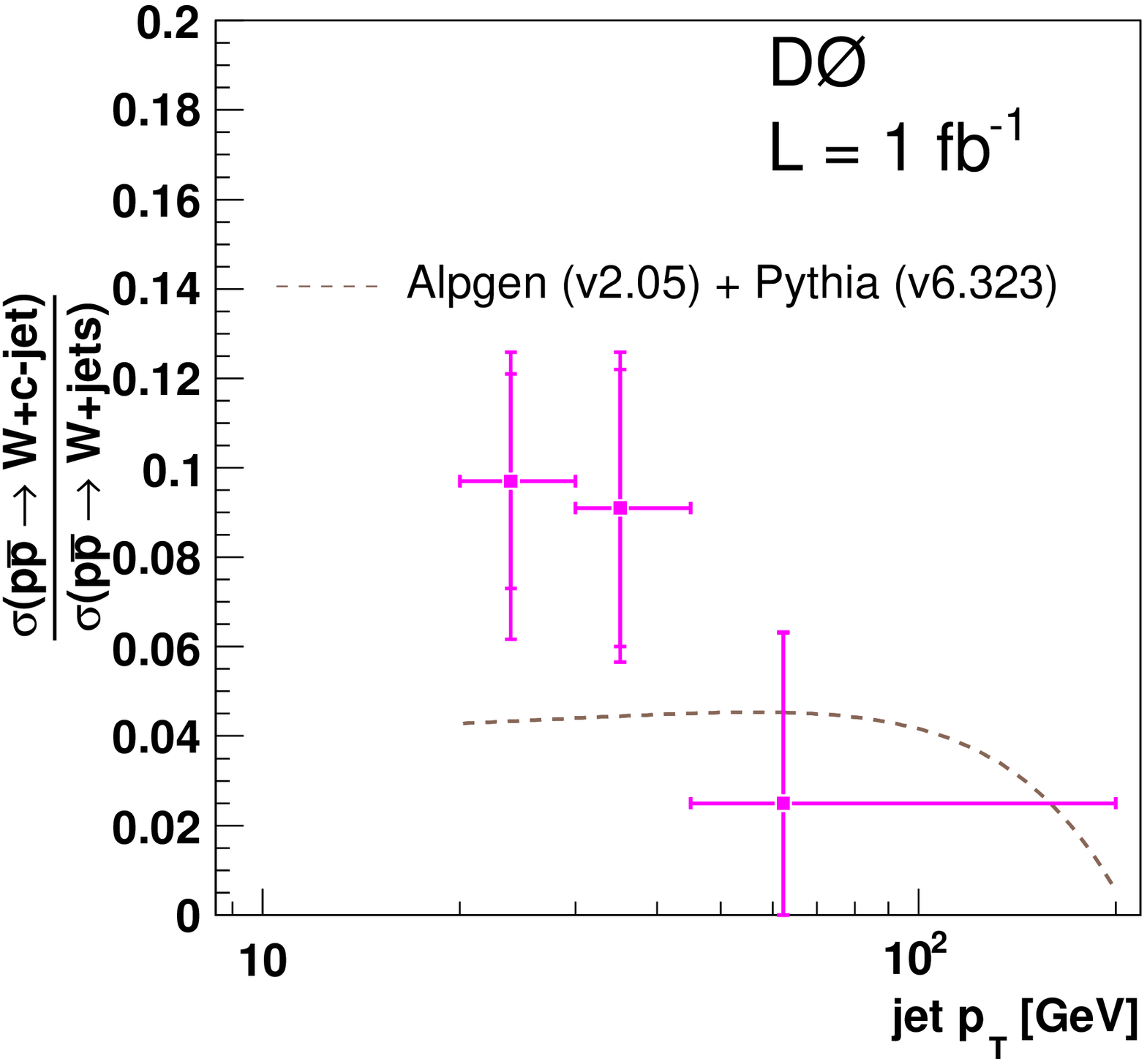}
 \caption{\label{fig:d0wc}
 Measured ratio 
% $[\sigma\left(\W+c\mathrm{-jet}\right) / \W+\mathrm{jets}]$
% $[\sigma\left(\wpcj\right) / \wpj]$
 $[\sigma\left(\wpcj\right) / \W+\mathrm{jets}]$        % to allow a line break
 from ref~\cite{ref:d0wc}. }
\end{minipage}
\end{figure}

Both collaborations have preliminary results from measurements
of \wpbj\ production.
The \DZ\ collaboration set a limit of
$\sigma\left(\ppbar\to\W b\bar{b}\right) < 4.6\un{pb}$ at 95\% C.L.
using $382\pbo$ of data~\cite{ref:d0wbb}. 
The jets' \pt\ was required to be
above 20\GeV\ and their direction to satisfy $\left|\etajet\right|<2.0$,
and only events with one or two jets were used.
%The $b$ jets were also required to have an angular separation of
%$\Delta R > 0.75$.
On the first day of the conference, the \CDF\ collaboration 
released preliminary results from a measurement	of the
$b$-jet production cross section in association with a \W\
boson: 
$\sigma_{b - \mathrm{jets}}\left(\ppbar\to\W + b-{\mathrm{jets}}\right)
\cdot B\left(\W\to l\nu\right) = 2.74\pm0.27\stat\pm0.42\syst\un{pb}$.
The dataset used in this measurement had an integrated luminosity
of $1.9\pbo$~\cite{ref:cdfwb} (see fig~\ref{fig:cdfwb}).
Jets were reconstructed with $R_{\mathrm{cone}}=0.4$, and counted as
$b$ jets if $\Delta R_{bj} < 0.4$, $E_T^{\mathrm{jet}} > 20\GeV$,
and  $\left|\eta^{\mathrm{jet}}\right| < 2$.
The measured cross section is significantly higher than 
the \alpgen\ prediction of $0.78\un{pb}$.

Both collaborations studied the rate of \wpcj\ production.
The \CDF\ collaboration measured 
$\sigma\left(\ppbar\to\wpcj\right)
\cdot B\left(\W\to l\nu\right) = 9.8\pm2.8\stat^{+1.4}_{-1.6}\syst\pm0.6\lumi\un{pb}$
using $1.8\fbo$ of data~\cite{ref:cdfwc}. 
The $c$-jet \pt\ was required to be above 8\GeV\ and 
their direction to satisfy $\left|\eta\right|<3.0$.
A recent preliminary result from the \DZ\ collaboration
was shown at the conference, they measure the ratio
\begin{equation}
R = \frac {\sigma\left(\ppbar\to\wpcj\right)}
          {\sigma\left(\ppbar\to\wpj\right)},
\end{equation}
and find $R=\left(7.4\pm1.9\stat^{+1.2}_{-1.4}\syst\right)\%$
using $1\fbo$ of data~\cite{ref:d0wc} (see also fig~\ref{fig:d0wc}).
The jets' \pt\ was required to be above 20\GeV\ and 
their direction to satisfy $\left|\etajet\right|<2.5$.
The measured fraction is higher than
the \alpgen\ prediction of $\left(4.4\pm0.3\pdfs\right)\%$.

Finally, the \CDF\ collaboration measured the differential 
\wpj\ production cross section as a function of the 
number of jets and the jet transverse energy using
$320\pbo$ of data~\cite{ref:cdfwj}. 
Jets are required to have  $\left|\eta\right|<2.0$.
The measured cross sections are compared to next-to-leading
order predictions and to predictions from two matched
MC generators.

\section {Modeling \boldmath \zpj\ production as a background}

\zpj\ production appears at a lower rate than \wpj\ production,
but has much less background, making it a good process
for tuning the simulations.
Usually it suffices to normalize simulated cross sections 
according to cross sections calculated at next-to-leading order (NLO) by the MCFM 
program~\cite{ref:mcfm}, though next-to-next-to-leading 
order calculations are also used sometimes.
As noted above, the strong dependence of the cross sections
on the kinematic cuts must be taken into account.
Some analyses normalize the total rate to data, 
for example, ref~\cite{ref:fcnc} where the apparent
data vs. MC discrepancy for \W\ plus a few jets production can be
resolved either by jet energy calibration effects or
by the appropriate choice of the hadronization and factorization
scales.

\begin{figure}
\centering
\includegraphics[width=\halfWid\linewidth]{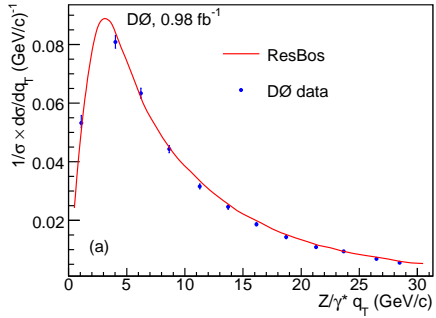}
\caption{\label{fig:zpt}
Normalized differential cross section as a function of
transverse momentum for the inclusive sample in ref~\cite{ref:d0resbos}.
}
\end{figure}

The kinematics of \zpj\ production can also be tuned to data.
Recently the \DZ\ collaboration noted that \resbos~\cite{ref:resbos}
calculations match their observed $d\sigma / d\pt^{\Z}$ distributions 
well~\cite{ref:d0resbos} (see fig~\ref{fig:zpt}),
and are starting to use \resbos\ as a surrogate	to the data,
reweighting \alpgen+\pythia\ MC so it agrees with the
\pt\ spectrum predicted by \resbos. This reweighting
is also carried over to \wpj\ production.
During the conference, \alpgen\ authors commented
that this may be due to the tuning of \alpgen\ parameters
used at \DZ, as \alpgen\
with the default parameters agrees with \resbos~\cite{ref:priv}.

The \DZ\ collaboration also compared differential \zpj\ cross sections
between data and the predictions of the \sherpa~\cite{ref:sherpa} 
and \pythia\ event generators~\cite{ref:d0sherpa}. As expected,
since \pythia\ is a parton shower generator it does not generate
sufficient additional radiation, while \sherpa\ simulates these
aspects adequately. Some inaccuracies are also evident
in the \pythia\ simulation of the unsigned rapidity difference between the 
two leading jets. It is interesting to note that again, 
\sherpa\ simulates the distribution adequately.

The differential \zpj\ cross sections were also measured
by the \CDF\ collaboration, which compared
the data both to NLO calculations (performed using MCFM) and to
different matched MCs~\cite{ref:cdfzpj}.
They found excellent agreement between the data and the  
NLO calculations of the cross sections as a function of
$N_{\mathrm{jets}}$ and $|y^{\mathrm{jet}}|$.

\section {Modeling \boldmath \wpj\ production as background}

Top quark measurements by the \CDF\ and \DZ\ collaborations
model \wpj\ production on the basis of the differential
distributions predicted by matched \alpgen\ MC. 
There are some indications that small corrections to
the differential distributions may be required, and these
are treated as systematic uncertainties in some analysis (e.g.
ref~\cite{ref:cdfst}). 
On the other hand, there is a clear need for
correcting the predicted integrated \wpj\ and \whf\ 
cross sections, and these are normalized to data,
after other backgrounds (multijets, dibosons, etc.) are subtracted.
Typically, \wpj\ production is normalized to data before
$b$ tagging, and the fraction of \whf\ in the total \wpj\
production is then fitted to data after $b$ tagging.

\begin{figure}
\centering
\includegraphics[width=0.5\linewidth]{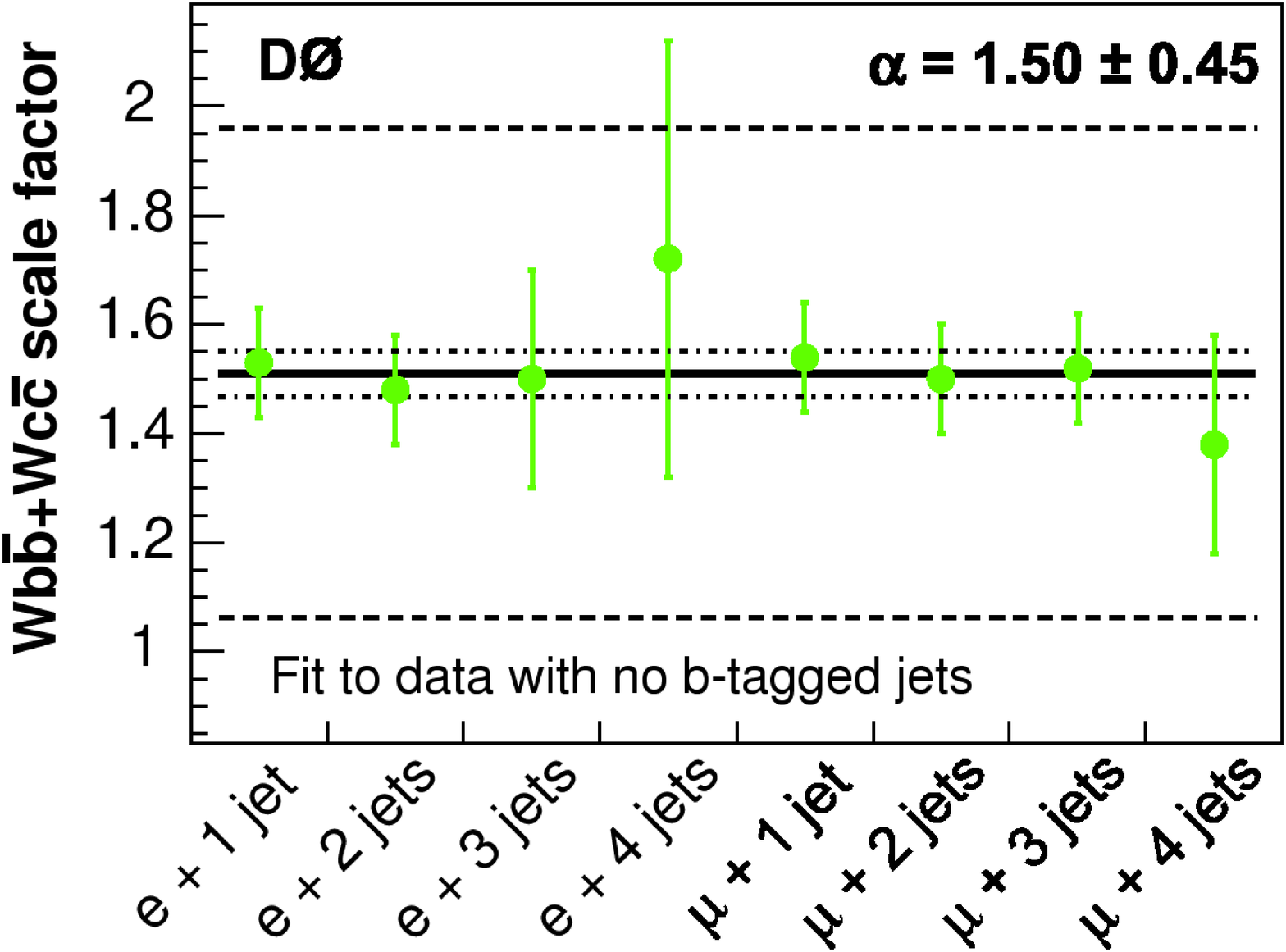}
\caption{\label{fig:khf}
Measurements of \khf\ from ref~\cite{ref:d0st}.
The points are the measured correction factor in each dataset. 
The solid line is the average of these values. 
The dot-dash inner band shows the uncertainty from the the fit
to the eight data points.
The dashed outer line shows the uncertainty  used in
the analysis.
}
\end{figure}

In \DZ\ analyses the \wpj\ normalization differs from analysis
to analysis.
It is determined either by counting events with one or two jets,
or by fitting a discriminant in \ttbar\ signal samples (with $\geq3$ jets).
The fraction of heavy flavor in the \wpj\ was normalized
to data using the number of events with no $b$ tagged jets~\cite{ref:d0st}.
This yielded a correction of $\khf=1.5\pm0.45$ (see fig~\ref{fig:khf})
to be applied to the heavy flavor fraction simulated by \alpgen. 
Later analyses used tighter selection cuts and normalized
the fraction of events with no $b$ tags (rather then their
absolute number). Tests for systematic effects revealed that
this factor is sensitive to the other background in these samples,
and to the jet selection. The resulting normalization was 
$\khf=1.17\pm0.18$. Oddly, switching from \alpgen\ version
2.05 to version 2.12 changed the \whf\ cross section by a 
factor of $\approx 2$, which together with more minor
improvements to the analyses yielded a new value of $\khf=1.9\pm0.3$

In \CDF\ analyses, distributions of jet-flavor discriminating
variable such as the output of neural network $b$ tagger or the
mass of reconstructed secondary vertex in $\W+1\mathrm{jet}$ data
are fit to the sum of light, charm, and bottom jet templates.
This yields $\khf=1.4\pm0.4$.
The \wlf\ component of the $b$-tagged samples is 
then determined 
by applying $b$-tagging rates either to 
the data before $b$ tagging or to \wlf\ MC samples.
%by subtracting other backgrounds including
%\whf\ from the data before $b$ tagging, and applying
%$b$-tagging rates.
To date, $K_{\bbbar} = K_{\ccbar} = 1$ and $K_c = 1$ are used
in all Tevatron top quark measurements as they are consistent
with the data.

In searches for physics beyond the SM in top samples,
the data often allows for significant non-standard production.
When the \vpj\ background is normalized to data
in the signal samples (e.g. $\geq3$ jets for \ttbar),
the possible non-standard production can affect the 
measured normalization. For example,
in the searches for resonant top-pair production
this is explicitly accounted for~\cite{ref:mtt}.

\section {MC use in the modeling of multijet background}
\label{sec:multijet}

Background from multijet production with a fake lepton
is modeled using various data-driven techniques with little
use of MC inputs.
Still, there is a place for MC in the modeling of
multijet background.
The data samples on which these estimations are based
are typically dominated by three jet events that
are reconstructed as a lepton and two jets.
As three jet production can be easily generated by MC
techniques, such samples can used to verify that 
the data driven techniques work as intended.

\section {\DZ\ experience in using matched MC}

Most of Tevatron experience with using matched MC is with \alpgen,
as at the time it was the only matched MC that could be run, 
integrated with the experiments' software, and be mass produced.
Both experiments produced a wide range of physics results
using \alpgen. But this success did not come without some
difficulties, and a few lessons may be learned from \DZ's
experience.

The \DZ\ collaboration overlays data collected with no trigger bias
over the simulated hard scatters, to simulate additional interactions,
pileup effects, noise, etc.
As the Tevatron luminosity increases, it is desirable to overlay both
older data and the very latest data.
But data quality issues can arise at the late stages of data analysis,
and data that was overlaid over the MC may later
be classified as bad. Thus \DZ\ removes events from the MC samples 
if their overlaid data was of bad quality. 
The HF removal described in sec~\ref{sec:match} is
also performed in this post-processing step.

This contributes to the problem of long turnaround times.
Once a new feature is put into the MC, it waits for the MC
authors to make a software release, then the experiment needs
to build and verify its software using the new MC version, 
the samples need to be produces (lots of events needed with
one or no extra jets, generating events with many extra jets
is slow), the post-production described above is done, and
finally the new samples must be propagated through the
physics analyses. Overall, six to twelve months pass before
a change in the MC is evaluated. The long turnaround times
have made even small mistakes, such as in setting random seeds,
very costly. This limits our ability to generate 
sufficient samples to study systematics.

When using matched MC, the different parton-jet bins must
be matched with the correct weights. These weights have a
wide range, which complicates the statistical analysis
of the simulated background. This wide range is unavoidable when 
simulating extra jet production, as more detailed simulation 
of the rare processes with many extra jets is needed.
The weights are also sample dependent, and so depend also
on the post-processing described above.
Therefore the simulated samples must be frozen, 
resulting in difficult book keeping which
is further complicated by the need for generating \zpj\ MC in different
mass bins. A possible lesson is that MC production should be
designed to avoid any post processing that changes the 
matching weights. E.g. in order to avoid changes due to data quality,
it may be possible to overlay the same set of data
events on top of all MC samples to be matched together.

\section {Conclusions}

Modeling \wpj\ and \zpj\ backgrounds purely from the simulation
is insufficient, and additional inputs from data are required.
Though a generic solution can work for most analyses, some analyses
can make due without the most sophisticated treatments, and 
some (especially new physics searches) have their own unique 
requirements. Several approaches are used to estimate the heavy
flavor contributions, and the overall \wpj\ contributions.
The data indicates that \wpj\ and in particular \whf\ production 
is more copious than predicted by \alpgen.

Matched \alpgen\ MC has been used extensively for the last couple
of years and was able to meet all our physics needs.
Some possible inaccuracies have been identified, in particular
in jet angular variables, and some technical lessons can be
learned. Other generators seem promising, but have received much
less scrutiny at the Tevatron.

%-----------------------------------------------------------------------

\end{document}